# Design, production and characterization of mirrors for ultra-broadband, high-finesse enhancement cavities

M. Trubetskov,[1,*] T. Amotchkina,[1] N. Lilienfein,[1,2] S. Holzberger,[1,2]
F. Krausz,[1,2] I. Pupeza,[1,2] V. Pervak[2]

[1]*Max-Planck-Institut für Quantenoptik, Hans-Kopfermann-Str. 1, Garching 85748, Germany*
[2]*Ludwig-Maximilians-Universität München, Am Coulombwall 1, Garching 85748, Germany*
*\*Michael.Trubetskov@mpq.mpg.de*

**Abstract:** To enable the enhancement of few-cycle pulses in high-finesse passive optical resonators, a novel complementary-phase approach is considered for the resonator mirrors. The design challenges and first experimental results are presented.

## 1. Introduction

Passive optical resonators or *enhancement cavities* (EC) serve manifold applications in linear as well as nonlinear optics. Early on, the ability to efficiently couple light to transverse eigenmodes of stable optical resonators has been used to enhance the interaction of light with a diluted gas sample and, thus, increase the sensitivity of linear absorption spectroscopy [1]. A recent example for the use of ECs in the field of nonlinear optics is high-order harmonic generation (HHG) of ultrashort near-infrared pulses at repetition rates of several tens of MHz [2]. Here, the pulses of a mode-locked laser are coherently stacked inside of a high-finesse EC, allowing for a pulse energy enhancement of a few orders of magnitude with respect to the original pulse train, which enables the intensities necessary to drive HHG in an intracavity gas target.

However, the same properties that underlie the sensitivity enhancement in EC-based spectroscopy also constitute a technical limitation of the optical bandwidth supported by a high-finesse EC; the spectral phase of an ultrashort pulse circulating inside of an EC in the steady state is affected by the single-roundtrip linear and nonlinear dispersion, enhanced by a factor close to the energy enhancement [3], resulting in a suboptimal interference at the input coupler of the cavity. Currently, the shortest pulse durations reported for high-finesse ECs are in the range 25-30 fs at central wavelengths around 1 μm [4]. Advancing multi-layer mirrors for high-finesse ECs to bandwidths supporting few-cycle pulses would allow for a quantum leap in the applications mentioned above. For instance, driving HHG in an EC supporting few-cycle pulses would enable the generation of powerful, isolated attosecond pulses at multi-MHz repetition rates, and, controlling the spectral phase of broadband mirrors with a high precision would allow for an efficient coupling to the cavity accounting for the intracavity nonlinearity [3]. Furthermore, increasing the bandwidth over which a high mirror reflectivity can be combined with a flat spectral phase would tremendously extend the possibilities of frequency-comb-based absorption measurements, allowing for massively parallel high-precision measurements.

In this contribution, the challenging design, production and characterization problem of dispersive mirrors (DM) providing a group delay dispersion (GDD) close to zero for a roundtrip in the EC and supporting an ultra-broadband spectral range is addressed. The novelty of this work is the fact that the complementary-pair approach to reach a 0-rountrip GDD in an ultra-broadband spectral range is applied.

## 2. Design approach

We consider the design problem of high-reflection mirrors for an EC working with *p*-polarized light and supporting circulating pulses of 15 fs at a finesse of 625. The target *p*-polarized reflectance $R^{(p)}$ is 100% and the target GDD is zero in the spectral range of interest from 910 nm to 1190 nm, at an angle of incidence of 1.5°. A challenge of this design problem is connected with the fact that such a broad bandwidth in combination with a GDD close to 0 cannot be supported by a single mirror. For the specified bandwidth single-mirror solutions provide unacceptably high GDD residual oscillations leading to destruction of circulating pulse after several dozens of bounces. Therefore, to design dispersive mirrors with the spectral characteristics specified above, a complimentary pair approach has been considered [5]. Two dispersive coatings, further referred to as *DM-Minus* and *DM-Plus,* have been designed.

In these designs, $Nb_2O_5$ as the high index materials and $SiO_2$ as a low index material were used; the substrate material was Suprasil and substrate thickness 6.35 mm. Refractive index wavelength dependencies of thin-film materials and substrates are described by a well-known Cauchy formula:

$$n(\lambda) = A_0 + A_1(\lambda_0/\lambda)^2 + A_2(\lambda_0/\lambda)^4, \tag{1}$$

where $A_0, A_1, A_2$ are dimensionless parameters, $\lambda_0 = 1000$ nm, $\lambda$ is specified in nanometers. The values of the Cauchy parameters of thin film materials and substrates are presented in Table 1.

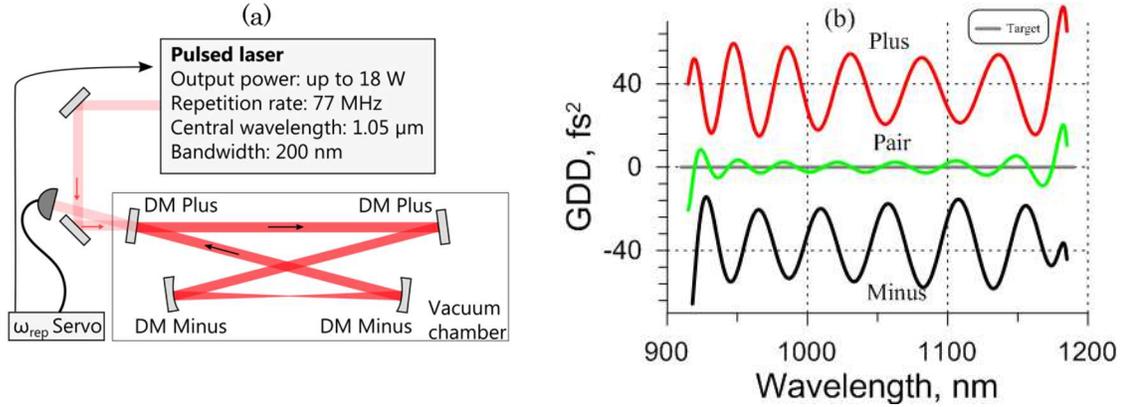

Fig. 1. (a): Schematic view of an enhancement cavity comprising three highly reflective mirrors(HR) and one input coupling mirror (IC), seeded by a broadband pulsed laser; (b): Theoretical GDD spectral dependencies

Table 1. Cauchy parameters of thin-film materials and substrates

| Material | $A_0$ | $A_1$ | $A_2$ |
| --- | --- | --- | --- |
| $Nb_2O_5$ | 2.218485 | 0.021827 | $3.99968 \times 10^{-3}$ |
| $SiO_2$ | 1.460472 | 0 | $4.9867 \times 10^{-4}$ |
| Suprasil | 1.443268 | $4.06 \times 10^{-3}$ | $6.9481764 \times 10^{-6}$ |

For the design purpose, the needle optimization technique with manual control incorporated into OptiLayer Thin Film Software were used [6]. Design *DM-Minus* consists of 66 layers and has total physical thickness of 10289 nm; design *DM-Plus* contains 64 layers and has physical thickness of 10207 nm. The reflectance values of each mirror are higher than 99.9% in the spectral range of interest. The GDD spectral dependencies of the designed mirrors are plotted in Fig. 1(b). The DM pair reflectance and GDD are calculated as:

$$R_{Pair}^{(p)}(\lambda) = \sqrt{R_{plus}^{(p)}(\lambda) \cdot R_{minus}^{(p)}(\lambda)}, \quad GDD_{Pair}^{(p)}(\lambda) = \left(GDD_{plus}^{(p)}(\lambda) + GDD_{minus}^{(p)}(\lambda)\right)/2 \quad (2)$$

Pair GDD is shown in Fig. 1(b): it is seen that oscillations of GDD can be suppressed if $GDD_{plus}$ and $GDD_{minus}$ are almost in antiphase.

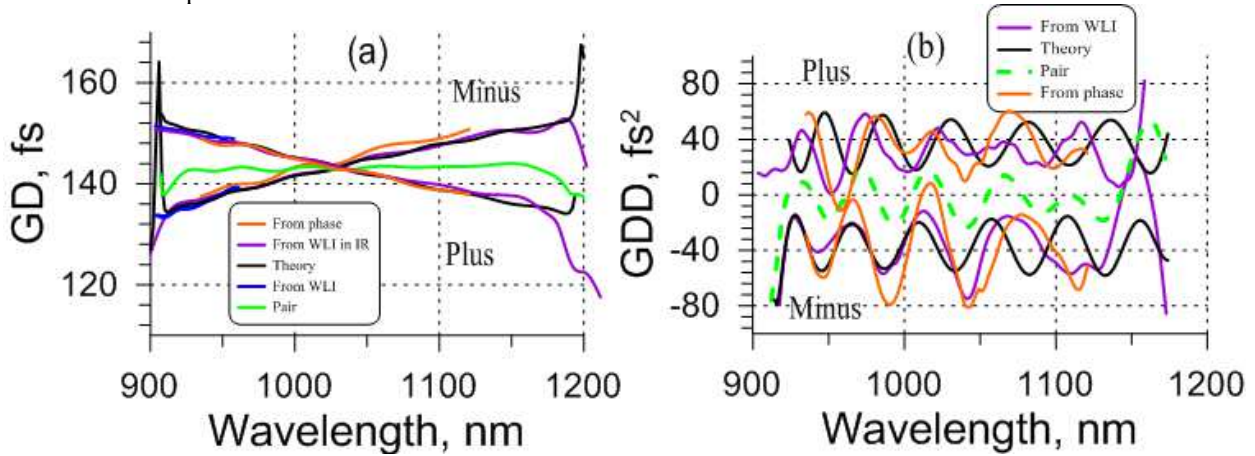

Fig. 2. Comparison of theoretical and experimental GD (a) and GDD (b). Pair GD and GDD are calculated from the WLI data.

## 3. Experimental results

The production of complementary pairs of DMs is extremely challenging, since it is necessary to deposit precisely matching pair with extremely high accuracy. The designed DMs were produced using Leybold Optics magnetron sputtering Helios plant, layer thicknesses were controlled with the help of well-calibrated time-monitoring [5]. The spectral transmittance of the produced DMs was measured at PerkinElmer Lambda 950 spectrophotometer and excellent correspondence between theoretical and experimental transmittance data has been achieved. Most critical for the application in ECs, however, are the measured phase characteristics of the DMs.

GD and GDD of the DM samples were extracted from the measurements provided by a white light interferometer (WLI) and processed with a specially developed algorithm [7]. WLI data was taken in two spectral ranges using two different detectors, namely in the range from 870 nm to 1100 nm and in the range from 900 nm to 1200 nm. In Fig. 2(a) the measured and theoretical GD of DMs are compared. It is seen that there exists a good correspondence between experimental and theoretical GD data as well as an excellent consistency between the GD data from two spectral ranges mentioned above. Pair GD calculated by Eq. (2) is also presented in Fig. 2(a) and it is seen that it exhibits a close to constant wavelength dependence. Additionally, the GD and GDD calculated from phase data, which was obtained by spatial-spectral interferometry with a coherent, nonlinearly-broadened pulsed laser source, are presented. The interferometer used in these measurements includes 24 passes on two sample mirrors, resulting in a measurement precision of about 5 mrad.

Fig. 2(b) represents a comparison of the experimental and theoretical GDD spectral dependencies of the produced DMs. The values of the pair GDD shown by green curve in Fig. 2(b) do not exceed 20 $fs^2$, the theoretical values of the pair GDD are in the range of 8-10 $fs^2$ (Fig. 2(b)). Simulations using the experimental data show that the produced mirrors allow the enhancement of 22-fs pulses at a cavity finesse of about 300. In the next deposition runs, the slight shift of the spectral characteristics of *DM-Plus* to the shorter wavelengths will be taken into account and performance of the complementary pair of DMs will be improved.

## 4. Conclusions

In conclusion, the problem of design, production and characterization of ultra-broadband, high-reflectivity dispersive mirrors for enhancement cavities has been considered. The novelty of the approach is using a complementary pair design which can, in principle, provide an unparalleled wide bandwidth for enhancement cavity mirrors, supporting circulating pulses of 15 fs duration. This approach is extremely challenging, since mirror pairs need to exactly match each other. Careful characterization of the results is required in order to achieve acceptable results. The authors believe that including feedback information obtained with the characterization techniques described here in the design and production processes constitutes a viable route towards high-performance mirrors for enhancement cavities supporting few-cycle pulses.

## 5. Acknowledgements

This work was supported by the DFG Cluster of Excellence, "Munich Centre for Advanced Photonics," (http://www.munich-photonics.de). T. Amotchkina has received funding from the European Union's Horizon 2020 research and innovation programme under the Marie Skłodowska-Curie agreement No 657596.

**References**

[1] B. A. Paldus and A. A. Kachanov, "An historical overview of cavity-enhanced methods," Can. J. Phys. **83**, 975–999 (2005).
[2] I. Pupeza, S. Holzberger, T. Eidam, H. Carstens, D. Esser, J. Weitenberg, P. Russbueldt, J. Rauschenberger, J. Limpert, Th. Udem, A. Tuennermann, T.W. Haensch, A. Apolonski, F. Krausz, E. Fill, "Compact high-repetition-rate source of coherent 100 eV radiation," Nature Photonics **7**, 608 (2013).
[3] S. Holzberger, N. Lilienfein, H. Carstens, T. Saule, F. Luecking, M. Trubetskov, V. Pervak, T. Eidam, J. Limpert, E. Fill, F. Krausz, I. Pupeza, "Femtosecond enhancement cavities in the nonlinear regime," Physical Review Letters **115**, 023902 (2015).
[4] S. Holzberger, N. Lilienfein, M. Trubetskov, H. Carstens, F. Luecking, V. Pervak, F. Krausz, I. Pupeza, "Enhancement cavities for zero-offset-frequency pulse trains," Optics Letters 40, 2165-2168 (2015).
[5] V. Pervak, A. V. Tikhonravov, M. K. Trubetskov, S. Naumov, F. Krausz, and A. Apolonski, "1.5-octave chirped mirror for pulse compression down to sub-3 fs," Appl Phys B **87**, 5–12 (2007).
[6] A. Tikhonravov and M. Trubetskov, OptiLayer Thin Film Software, www.optilayer.com
[7] T. V. Amotchkina, A. V. Tikhonravov, M. K. Trubetskov, D. Grupe, A. Apolonski, and V. Pervak, "Measurement of group delay of dispersive mirrors with white-light interferometer," Appl Opt **48**, 949–956 (2009).